\documentclass[3p,times]{elsarticle}

\usepackage{ecrc}
\usepackage{amssymb,amsmath}

\volume{00}

\firstpage{1}

\journalname{Nuclear Physics A}

\runauth{Tetsuo Hyodo}

\jid{NUPHA}

\jnltitlelogo{Nuclear Physics A}

\CopyrightLine{2012}{Published by Elsevier Ltd.}

\usepackage{amssymb}
\usepackage[figuresright]{rotating}

\newcommand{\chiral}{SU(3)$_R\, \times \, $SU(3)$_L$\,}

\newcommand{\ket}[1]{| \, #1 \, \rangle}

\begin{document}

\begin{frontmatter}

\dochead{}



\title{Recent developments in antikaon--nucleon dynamics}

\author{Tetsuo Hyodo}
%\author[b]{Author2}\author[a,b]{Author3}

\address{Department of Physics, Tokyo Institute of Technology, Meguro 152-8551, Japan}

%\address[b]{Institution 2, Country2}

\begin{abstract}
Stimulated by various experimental achievements, the study of $\bar{K}N$ dynamics now enters a new phase. The two-body $\bar{K}N$ interaction is largely constrained by recent experimental data, and the nature of the $\Lambda(1405)$ resonance is being unveiled by several theoretical analyses. These findings provide a basic tool for applications to $\bar{K}$-nuclear systems. We summarize the current status of the $\bar{K}N$ phenomenology and outline the future direction in this field.

\end{abstract}


\end{frontmatter}

% \linenumbers

\section{Introduction}
\label{sec:intro}
%%%%%%%%%%%%%%%%%%%%%%%%%%%%%%%%%%%%%%%%%%%%%%%%%%%%

% history and present
In 1959, Dalitz and Tuan predicted the existence of a $\pi Y$ resonance state just below the $\bar{K}N$ threshold, based on the analysis of the $\bar{K}N$ scattering lengths~\cite{Dalitz:1959dn,Dalitz:1960du}. Within two years, the resonance, which is now called $\Lambda(1405)$, was observed experimentally in the invariant mass spectrum of the $\pi\Sigma$ system~\cite{Alston:1961zz}. Shortly after the finding of $\Lambda(1405)$, by regarding it as a $\bar{K}N$ quasi-bound state, the possible existence of the $\bar{K}NN$ state (and in general $\bar{K}$-nuclei) was discussed~\cite{PL7.288}. It is remarkable that the essential ingredients of the recent $\bar{K}N$ phenomenology have already been considered about half a century ago. Nowadays QCD is established as the fundamental theory of strong interaction, and the complicated dynamics of the $\bar{K}N$ system together with the $\Lambda(1405)$ resonance can be understood by coupled-channel approach with chiral \chiral symmetry~\cite{Kaiser:1995eg,Oset:1998it,Oller:2000fj,Lutz:2001yb,Hyodo:2011ur}. This framework enables us to investigate the internal structure of $\Lambda(1405)$, and provides systematic improvement of the theoretical description in response to the accumulation of precise experimental data.

% Kbar N interest, Kbar nuclei
There are several reasons why the $\bar{K}N$ system with $\Lambda(1405)$ is interesting to be studied. From a theoretical point of view, (anti-)kaons represent the interplay between spontaneous and explicit chiral symmetry breaking with three flavors. Because of this, the antikaons possess various peculiar features, one of which is the strong attraction of the $\bar{K}N$ system in isospin $I=0$ channel. This strong attraction is the driving force to generate $\Lambda(1405)$ as a quasi-bound state of $\bar{K}N$ embedded in the $\pi\Sigma$ continuum. The quasi-bound picture of $\Lambda(1405)$ also attracts phenomenological interest, since it leads to the possibility of $\bar{K}$ bound states in nuclei~\cite{Akaishi:2002bg} which is one of the hot topics in recent strangeness nuclear physics. Although many theoretical and experimental efforts have been put on the $\bar{K}$-nuclei, the lack of precision of the experimental data hindered a reliable subthreshold extrapolation of the $\bar{K}N$ interaction~\cite{Hyodo:2007jq}.

% experiments
This situation has changed recently. A new accurate measurement of the 1s level of the kaonic hydrogen is performed by SIDDHARTA~\cite{Bazzi:2011zj}. The energy shift $\Delta E$ and the width $\Gamma$ of the kaonic hydrogen atom are related to the $K^{-}p$ scattering length, so the precise data provides a severe constraint on the elastic $K^{-}p$ scattering amplitude at the threshold. In addition, mass spectra of the $\pi\Sigma$ system have been measured in photoproduction processes by LEPS~\cite{Niiyama:2008rt} and CLAS~\cite{Moriya:2009mx}, and in $pp$ collisions by HADES~\cite{Agakishiev:2012xk}. Because the $\pi\Sigma$ spectrum is the only place where the $\Lambda(1405)$ resonance is seen as a peak structure, it is an important observable to extract the $\Lambda(1405)$ properties. At the same time, the $\pi\Sigma$ spectrum constrains the subthreshold $\bar{K}N$ amplitude through the coupled-channel effect. These new experimental results are summarized in Table~\ref{tab:exp}, some of which are also reported in this conference~\cite{Curceanu,Schumacher,Fabbietti}.

\begin{table}[tbp]	
\begin{center}
\caption{
Recent experimental results for the $\bar{K}N$ scattering and $\Lambda(1405)$. 
\label{tab:exp}
}
\begin{tabular}{lll}
\hline
Collaboration & Reaction & Observables \\ \hline
SIDDHARTA~\cite{Bazzi:2011zj,Curceanu} & stopped $K^{-}$ & ($\Delta E$, $\Gamma$) of kaonic hydrogen\\
LEPS~\cite{Niiyama:2008rt} & $\gamma p\to K^{+}\Lambda(1405)$ @ $E_{\gamma}=$1.5--2.4 GeV & $\pi^{\pm}\Sigma^{\mp}$ spectra, cross sections   \\ 
CLAS~\cite{Moriya:2009mx,Schumacher} & $\gamma p\to K^{+}\Lambda(1405)$ @ $E_{\gamma}=$1.56--3.83 GeV & $\pi^{\pm}\Sigma^{\mp}$ and $\pi^{0}\Sigma^{0}$ spectra, cross sections   \\ 
HADES~\cite{Agakishiev:2012xk,Fabbietti} & $pp\to K^{+}p\Lambda(1405)$ & $\pi^{\pm}\Sigma^{\mp}$ spectra, cross sections   \\ 
\hline
\end{tabular}
\end{center}
\end{table}

% summary
Here we report on recent developments in the study of the $\bar{K}N$ dynamics. We discuss how the $\bar{K}N$ scattering is described in chiral coupled-channel approach, and show that the structure of $\Lambda(1405)$ can be extracted in various methods in Section~\ref{sec:L1405}. An attempt to construct a realistic $\bar{K}N$ interaction with new threshold constraints is discussed in Section~\ref{sec:KNint}. Comparison of several chiral models is presented and applications to nuclear systems are also mentioned. Section~\ref{sec:conclusion} summarizes this paper and indicates future prospects of the $\bar{K}N$ phenomenology.

%%%%%%%%%%%%%%%%%%%%%%%%%%%%%%%%%%%%%%%%%%%%%%%%%%%%
\section{$\Lambda(1405)$ in $\bar{K}N$-$\pi\Sigma$ scattering}
\label{sec:L1405}
%%%%%%%%%%%%%%%%%%%%%%%%%%%%%%%%%%%%%%%%%%%%%%%%%%%%

In this section, we discuss the strangeness $S=-1$ meson--baryon scattering and the structure of the $\Lambda(1405)$ resonance from the viewpoint of chiral SU(3) dynamics. In Section~\ref{sec:MBscattering} we introduce basic concepts of the chiral coupled-channel approach for meson--baryon scattering~\cite{Kaiser:1995eg,Oset:1998it,Oller:2000fj,Lutz:2001yb,Hyodo:2011ur}. We then turn to the discussion on the internal structure of $\Lambda(1405)$ in Section~\ref{sec:structure}.

%------------------------------
\subsection{Meson--baryon scattering in chiral dynamics}
\label{sec:MBscattering}

Historically, the $S=-1$ meson--baryon scattering has been successfully described by a vector meson exchange model with flavor SU(3) symmetry~\cite{Dalitz:1967fp}. The reason behind the success of this model can be attributed to the chiral low energy theorem for the $s$-wave meson--baryon interaction by Tomozawa and Weinberg~\cite{Tomozawa:1966jm,Weinberg:1966kf}, which reproduces the same group theoretical structure with the vector meson exchange potential, with the help of the KSRF relation~\cite{Kawarabayashi:1966kd,Riazuddin:1966sw}. The low energy theorem is a model-independent consequence of spontaneous breaking of chiral \chiral symmetry. Moreover, the establishment of chiral perturbation theory~\cite{Weinberg:1979kz,Gasser:1985gg} enables a systematic construction of the interaction based on chiral counting rule. In this way, chiral coupled-channel approach~\cite{Kaiser:1995eg,Oset:1998it,Oller:2000fj,Lutz:2001yb,Hyodo:2011ur} has been developed by combining the low energy meson--baryon interaction and the dynamical framework of coupled-channel scattering.

The meson--baryon interaction $V$ can be systematically derived in chiral perturbation theory as
\begin{align}
   V
   =&
   V_{\text{TW}}+V_{\text{Born}}
   +V_{\text{NLO}}
   +\cdots ,
   \label{eq:interaction}
\end{align}
where the leading order contribution consists of Tomozawa--Weinberg ($V_{\text{TW}}$) and Born ($V_{\text{Born}}$) terms, the next-to-leading order (NLO) corrections are summarized in $V_{\text{NLO}}$, and the dots represent the $\mathcal{O}(p^{3})$ contributions. These interactions are matrices in meson--baryon channel space, and off-diagonal components represent transition among different channels. Except for the meson decay constants which are fixed by the dynamics of Nambu--Goldstone bosons, there is no free parameter in $V_{\text{TW}}$ and all the coupling strengths are determined by SU(3) symmetry. Born terms include the axial coupling constants $D$ and $F$ which are given by neutron and hyperon beta decays. In the NLO terms, we have seven low energy constants which are not constrained by chiral symmetry.

\begin{figure}[tb]
\begin{center}
  \includegraphics[width=0.65\textwidth,bb=0 0 700 100]{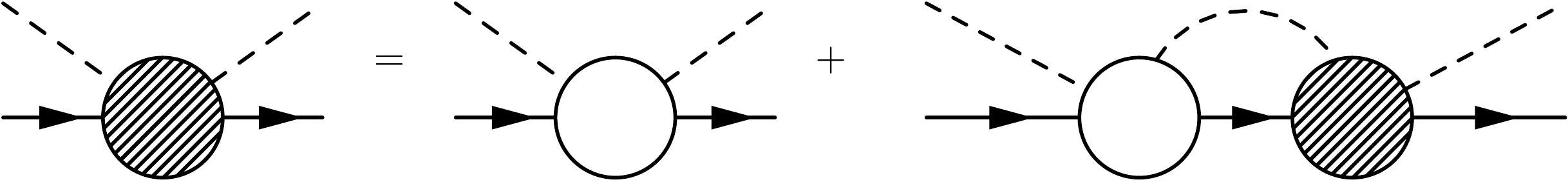}
\end{center}
\caption{Diagrammatic illustration of the Bethe--Salpeter equation for meson--baryon scattering. The shaded (empty) blob represents the scattering amplitude $T$ (interaction kernel $V$).}
\label{fig:BSE}
\end{figure}

In contrast to the two-flavor $\pi N$ scattering where the perturbative treatment works well, the two-body interaction in the strangeness sector is in some cases strong enough to generate a quasi-bound state. This is because the chiral low energy interaction has derivative coupling, so the leading interaction is proportional to the NG boson energy, which is not small for $\bar{K}$ due to the strange quark mass. Under such circumstances, it is mandatory to sum up the diagrams to full order by solving the scattering equation. The dynamical aspect of the coupled-channel meson--baryon scattering is taken care of by the Bethe--Salpeter equation (Fig.~\ref{fig:BSE}) with on-shell factorization, which leads to the scattering amplitude $T(W)$ with total energy $W$ as
\begin{align}
   T(W)
   =&\frac{1}{V(W)^{-1} -G(W;a)}
   \label{eq:amp} ,
\end{align}
where $G(W;a)$ is the loop function with subtraction constant $a$. This amplitude can also be derived in the N/D method~\cite{Oller:2000fj}. In this way, one obtains the scattering amplitude $T(W)$ which is consistent with chiral symmetry in the low energy region and satisfies unitarity in coupled channels. Moreover, the theoretical description can be systematically improved by introducing higher order corrections to the interaction kernel $V(W)$ based on chiral perturbation theory. In fact, the concept of the chiral coupled-channel approach is similar to the chiral effective field theory approach for the nuclear force~\cite{Machleidt:2011zz} where the $NN$ potential is derived from effective field theory which is then plugged into the scattering equation to obtain the full amplitude. 

%------------------------------
\subsection{Structure of the $\Lambda(1405)$ resonance}
\label{sec:structure}

The $\Lambda(1405)$ resonance is a negative parity excited baryon with strangeness $S=-1$, isospin $I=0$, and spin $J=1/2$. Its structure has been a long-standing problem in hadron spectroscopy, because the conventional quark model does not well reproduce the mass of $\Lambda(1405)$, in spite of the good description of other negative parity excited baryons~\cite{Isgur:1978xj}. One reasonable explanation of this deviation may be the meson--baryon molecule picture as implied by the success of the dynamical meson--baryon scattering models. It should be however noted that the ``three-quark state'' and ``meson--baryon molecule'' should mix with each other \textit{via} strong interaction, as far as they have same quantum numbers. Hence, one has to pin down the \textit{dominant} component of the ``wave function'' of $\Lambda(1405)$ which  may be expanded as
\begin{equation}
    \ket{\Lambda(1405)}
    =N_3\ket{qqq}
    +N_5\ket{qqqq\bar{q}}
    +N_{\text{MB}}\left(\ket{\text{M}}\otimes\ket{\text{B}}\right)
    +\cdots
    \label{eq:decomp} .
\end{equation}
One should be careful about the meaning of this decomposition, which contains subtle problems. First of all, the ``wave function'' of a resonance state is not a well-defined notion, and the decomposition of $\ket{qqqq\bar{q}}$ and $\ket{\text{M}}\otimes\ket{\text{B}}$ is not trivial. Thus, the definition of the internal structure of hadron resonances itself is an important issue to be investigated. In the following, we overview a series of activities to clarify the structure of $\Lambda(1405)$ in the chiral coupled-channel approach.

In the conventional hadron spectroscopy, the structure of excited hadrons has been identified by comparing several descriptions of different models. For instance, resonances which are well reproduced in chiral coupled-channel approach have been regarded as hadronic molecules. Ref.~\cite{Hyodo:2008xr} shows that this is not always the case. In general, the cutoff parameters in effective models should represent the effects that are not included in the model space of a given approach. Therefore, even if we start from the meson--baron scattering equation without explicitly introducing a bare resonance state, the generated resonance in the full amplitude can contain the three-quark or other components through the regularization parameter. In fact, by developing the natural renormalization scheme, Ref.~\cite{Hyodo:2008xr} shows that the $N(1535)$ resonance contains substantial non-meson--baryon contributions, while $\Lambda(1405)$ is dominated by the meson--baryon molecule component. To quantify the above discussion, the compositeness of dynamically generated states is introduced through the field renormalization constant~\cite{Hyodo:2011qc}. It is shown that the dynamically generated state in the natural renormalization scheme has large compositeness, which reinforces the conclusion of Ref.~\cite{Hyodo:2008xr}.

The quark structure of $\Lambda(1405)$ can be studied through the response to the variation of number of colors $N_{c}$~\cite{Hyodo:2007np,Roca:2008kr}, based on the method developed in the meson sector~\cite{Pelaez:2003dy}. The key point is that the $N_{c}$ dependence of the properties of $\bar{q}q$ mesons and $q^{N_{c}}$ baryons is model-independently known. Applying general $N_{c}$ scaling rules to hadron masses and coupling constants in chiral perturbation theory, we can trace the movement of the resonance pole along with the $N_{c}$ variation. In the baryon sector, it is important to include the $N_{c}$ dependence of the coupling strength of the leading order chiral interaction which originates in the modification of the flavor representation of baryons~\cite{Hyodo:2006yk,Hyodo:2006kg}. The pole trajectory is then compared with the $N_{c}$ scaling rule of the mass and width of the excited $N_{c}$-quark baryons, in order to judge whether the $N_{c}$-quark component is dominant or not. In the chiral coupled-channel approach, $\Lambda(1405)$ is accompanied by two poles~\cite{Jido:2003cb}, both of which do not follow the $N_{c}$ scaling rule of $N_{c}$-quark baryons. This indicates that, in the real world with $N_{c}=3$, $\Lambda(1405)$ is not dominated by the three-quark component.

A standard method to probe the structure of stable particles is the evaluation of electromagnetic properties. In Refs.~\cite{Sekihara:2008qk,Sekihara:2010uz}, the electromagnetic radii and form factors of $\Lambda(1405)$ are calculated by introducing an external photon current to the meson--baryon scattering amplitude. From the analysis of different probe currents, it is found that $\Lambda(1405)$ is dominated by the $\bar{K}N$ component, which distributes widely in spatial extent. On the other hand, the form factors obtained at the pole position become complex and the interpretation of the imaginary part is not straightforward. This issue is further pursued in Ref.~\cite{Sekihara:2012fu} where the spatial ``size'' of unstable states is defined using the finite volume effect. When a two-body bound state is put in a finite box with periodic boundary condition, the mass of the bound state is shifted from the value in the infinite volume. This mass shift is related to the spatial size of the bound state, so the size of a stable particle can be extracted from the volume dependence of the mass. This relation is generalized to closed-channel components of unstable resonances, so that we can obtain a real valued ``size'' of resonances. Applying this technique to $\Lambda(1405)$, we find the size of the $\bar{K}N$ component inside $\Lambda(1405)$ is 1.8--1.9 fm. This is substantially larger than the normal three-quark baryons of 1 fm, and consistent with the dominance of meson--baryon molecule picture.

In all cases studied, the $\Lambda(1405)$ resonance is found to be dominated by the meson--baryon molecule structure, with the $\bar{K}N$ channel being the major component. This indicates that $\Lambda(1405)$ can indeed be realized as the $\bar{K}N$ quasi-bound state, which provides the foundation of the $\bar{K}$ bound states in nuclei. We emphasize again that this is \textit{not} because of the good description of $\Lambda(1405)$ in dynamical scattering model, but because of the detailed analyses from various viewpoints.

%%%%%%%%%%%%%%%%%%%%%%%%%%%%%%%%%%%%%%%%%%%%%%%%%%%%
\section{Realistic $\bar{K}N$--$\pi\Sigma$ interaction with SIDDHARTA}
\label{sec:KNint}
%%%%%%%%%%%%%%%%%%%%%%%%%%%%%%%%%%%%%%%%%%%%%%%%%%%%

The basic ingredient for the study of the $\bar{K}$ dynamics with nuclei is the two-body $\bar{K}N$ interaction. As mentioned in the introduction, many new experimental results on the $\bar{K}N$ system are accumulated in recent years. We first summarize the current situation of experimental data for the $\bar{K}N$ interaction in Section~\ref{sec:exp}. We then discuss theoretical models which include SIDDHARTA constraint~\cite{Bazzi:2011zj} in Section~\ref{sec:NLO}. In Section~\ref{sec:applications}, we briefly mention the applications to the $\bar{K}$-nuclear systems.

%------------------------------
\subsection{Experimental database for the $\bar{K}N$ interaction}
\label{sec:exp}

In developing theoretical models of the $\bar{K}N$ scattering, the commonly used input has been the total cross sections of the $K^{-}p$ scattering into elastic and inelastic channels, and threshold branching ratios, $\gamma$, $R_{n}$ and $R_{c}$. The cross sections cover a wide energy region above the threshold, but the data points are accompanied by relatively large error bars. Threshold branching ratios, on the other hand, are accurately determined, but are given only at the $\bar{K}N$ threshold energy. A comprehensive analysis of the meson--baryon scattering with these data has been performed in Ref.~\cite{Borasoy:2006sr} which provides large uncertainties on the value of the $\bar{K}N$ scattering lengths. Thus, we need additional information other than the cross sections and the branching ratios to constrain the $\bar{K}N$ interaction.

The energy shift $\Delta E$ and the width $\Gamma$ of the kaonic hydrogen is related to the $K^{-}p$ scattering length $a(K^-p)$. By using the effective Lagrangian approach, these are related up to $\mathcal{O}(\alpha,m_{d}-m_{u})$ as~\cite{Meissner:2004jr}
\begin{align}
\Delta E - \textrm{i}\frac{\Gamma}{2} = -2\alpha^3\,\mu_r^2\,a(K^-p)\left[1+2\alpha\,\mu_r\,(1-\ln\alpha)\,a(K^-p)\right]
\label{eq:MRR}.
\end{align}
This means that the determination of the kaonic hydrogen atomic level gives a constraint on the $K^{-}p$ elastic scattering amplitude at threshold. Thus, the precise measurement of $\Delta E$ and $\Gamma$ by SIDDHARTA~\cite{Bazzi:2011zj} gives a significant impact on theoretical studies of the $\bar{K}N$ interaction, as we will discuss in Section~\ref{sec:NLO}. Note that Eq.~\eqref{eq:MRR} is a model-independent relation, but based on the expansions in terms of $\alpha$ and $m_{d}-m_{u}$. It is also possible to directly calculate $\Delta E$ and $\Gamma$ by solving the Schr\"odinger equation, when the off-shell dependence of the $\bar{K}N$ potential is assumed. The result is, however, dependent on the form of the chosen potential.

Before going to the theoretical analysis, we would like to comment on recent activities for the $\pi\Sigma$ spectra from LEPS~\cite{Niiyama:2008rt}, CLAS~\cite{Moriya:2009mx}, and HADES~\cite{Agakishiev:2012xk}. First of all, the $\pi\Sigma$ spectrum is related to the subthreshold amplitude of the $\bar{K}N$ scattering, which is important to the $\bar{K}N$ interaction in nuclear bound states. Moreover, these experiments provide $\pi\Sigma$ spectra in different charge states, which reflect the interference effect of different isospin components~\cite{Nacher:1998mi}. Therefore, the data of $\pi\Sigma$ spectra can give constraints on the $\bar{K}N$ amplitude. On the other hand, experimental $\pi\Sigma$ spectra are contaminated by background contributions, interference effects, and final state interactions, which depend on each reaction process. In order to extract the information of the meson--baryon scattering, therefore, it is mandatory to perform detailed analysis of the reaction mechanism for each experiment. At this moment, we should check the consistency of the model prediction with the experimental spectra.

There is another possible observable, the $\pi\Sigma$ scattering length. So far there is no experimental data on the $\pi\Sigma$ scattering, but this information may be accessible by the final state interaction of the weak decays of $\Lambda_{c}\to \pi\pi\Sigma$~\cite{Hyodo:2011js} and also by lattice QCD~\cite{Torok:2009dg,Ikeda:2011qm}. The scattering length is given by a normalized number at fixed energy at threshold, so it is easy to implement as a constraint on the scattering amplitude. In addition, it is shown that the structure of the $\Lambda(1405)$ resonance is sensitive to the value of the $\pi\Sigma$ scattering length~\cite{Ikeda:2011dx}. Thus, theoretical and experimental efforts to extract the $\pi\Sigma$ scattering length would help to reduce uncertainties in the subthreshold $\bar{K}N$ amplitude.
 
%------------------------------
\subsection{The $\bar{K}N$ amplitude with SIDDHARTA and comparison of different models}
\label{sec:NLO}

In order to assess the impact of the SIDDHARTA constraint, its effect should be systematically studied. Such an analysis has been performed in the chiral coupled-channel approach~\cite{Ikeda:2011pi,Ikeda:2012au} with the interaction kernel up to NLO level in Eq.~\eqref{eq:interaction}. We consider three models, TW, TWB, and NLO which include $V_{\text{TW}}$, $V_{\text{TW}}+V_{\text{Born}}$, and $V_{\text{TW}}+V_{\text{Born}}+V_{\text{NLO}}$ respectively, and fit them to experimental data of the $K^{-}$ total cross sections, threshold branching ratios, and SIDDHARTA measurement. We use Eq.~\eqref{eq:MRR} to relate the kaonic hydrogen measurement to the $K^{-}p$ scattering length. All models contain six channel-dependent subtraction constants which are free parameters. The NLO model have additional seven low energy constants to be adjusted to fit experimental data.

\begin{figure}[tb]
\centering
\begin{minipage}[t]{8cm}
\includegraphics[width=8cm,bb=0 0 1200 750]{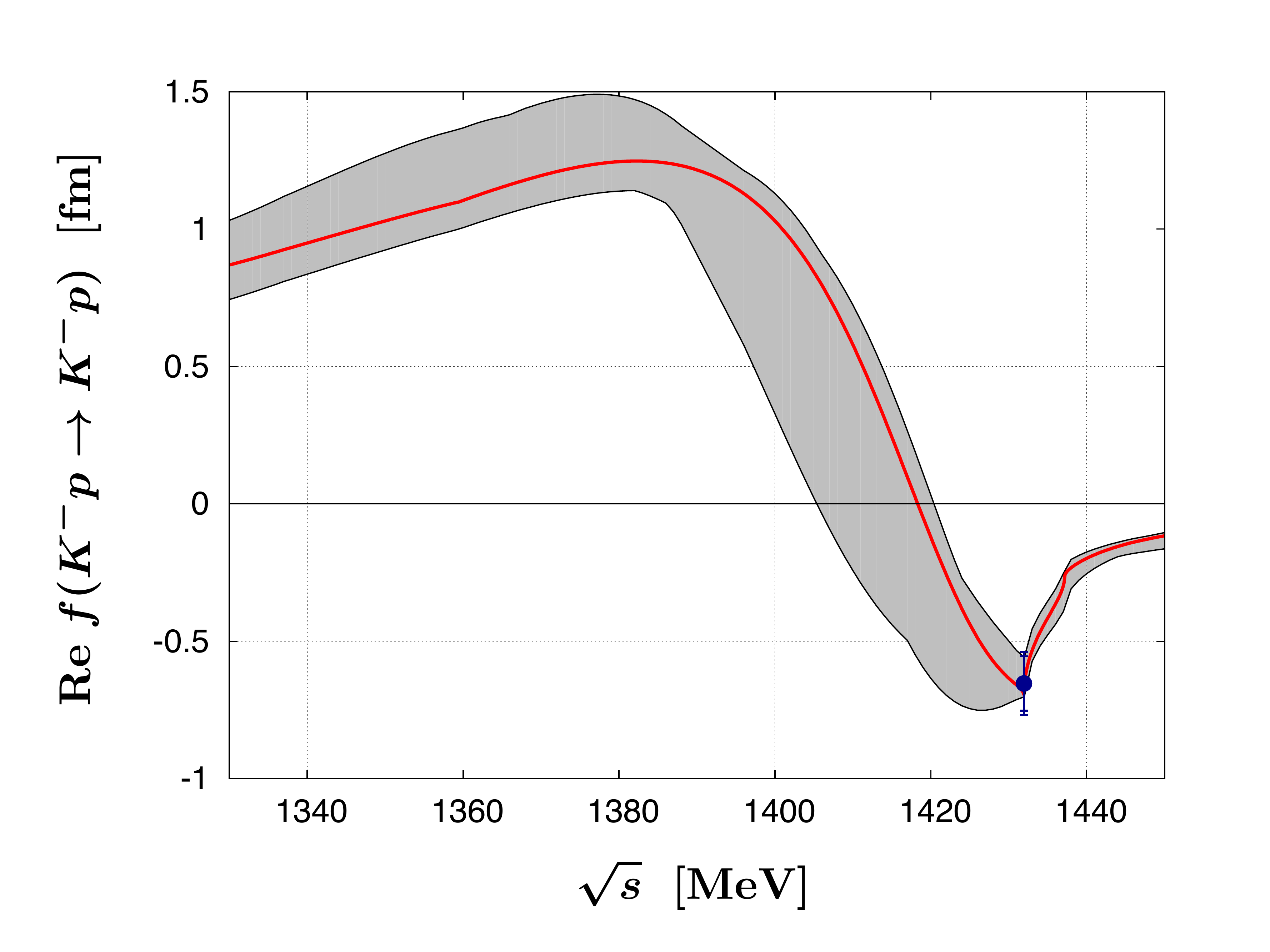}
\end{minipage}
\hspace{-1cm}
\begin{minipage}[t]{8cm}
\includegraphics[width=8cm,bb=0 0 1200 750]{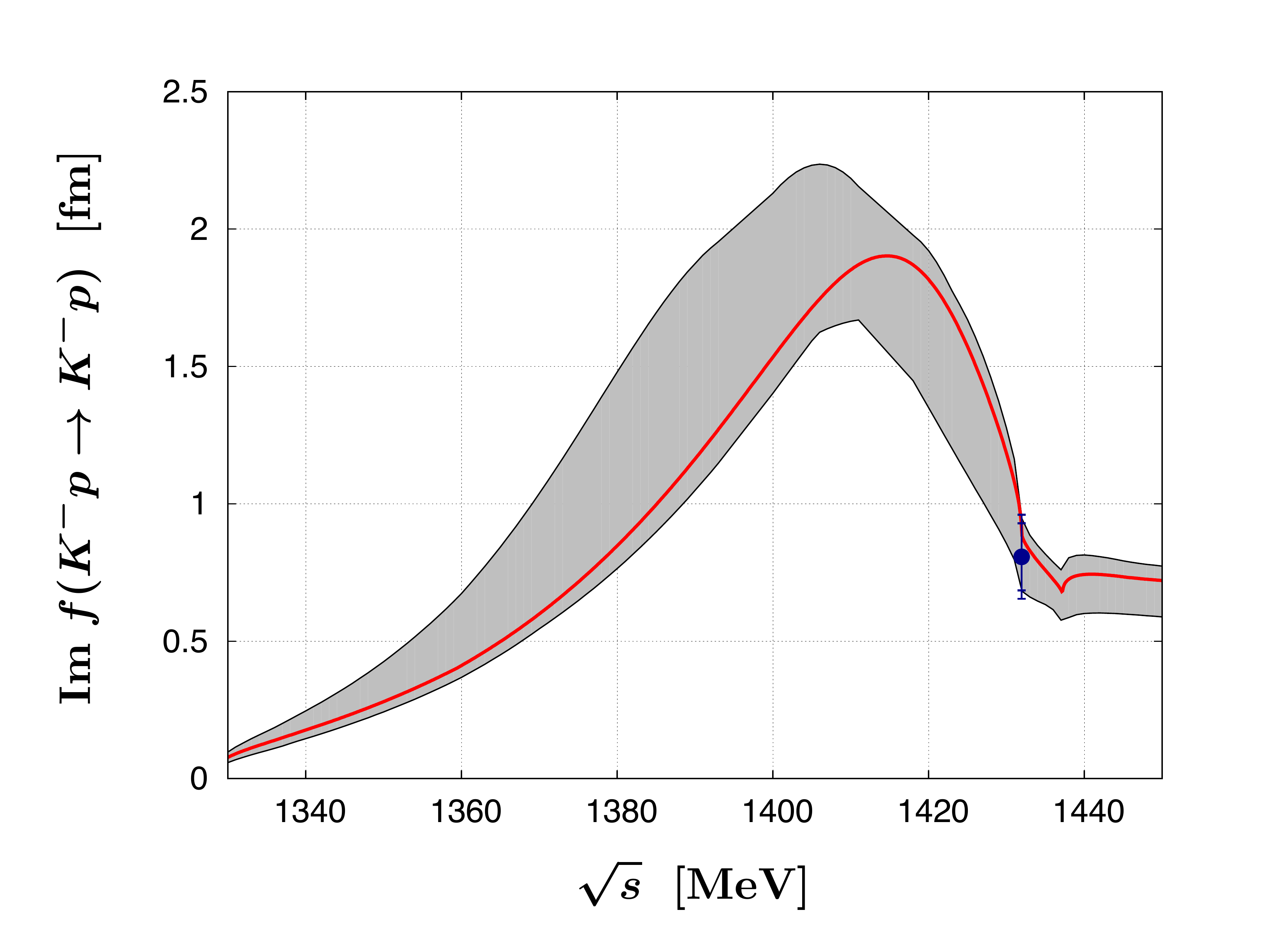}
\end{minipage}
\caption{Real part (left) and imaginary part (right) of the $K^-p \rightarrow K^-p$ forward scattering amplitude extrapolated to the subthreshold region. The SIDDHARTA constraints are indicated by the dots including statistical and systematic errors. The shaded bands represent theoretical uncertainties.} 
\label{fig:amplitude}
\end{figure}

As a result, we obtain reasonable descriptions of data in all three models. This means that the SIDDHARTA result of the kaonic hydrogen is fully consistent with the total cross sections and the threshold branching ratios. The calculated values of the kaonic hydrogen level shift $\Delta E$ and width $\Gamma$, together with the values of $\chi^{2}$/d.o.f., are summarized in Table~\ref{tab:IHW}. It is found that the global feature of experimental data can be reproduced by the leading order (TW and TWB) models with reasonable $\chi^{2}$/d.o.f.$\sim 1$. On the other hand, small overestimations of $\Delta E$ in TW and TWB models indicate that the NLO corrections are required to accommodate $\Delta E$ within the error bar of the SIDDHARTA measurement. In other words, the SIDDHARTA measurement is so accurate that the NLO terms are relevant. The accuracy of the SIDDHARTA measurement is also seen in subthreshold extrapolations. In Fig.~\ref{fig:amplitude}, we show the behavior of the $K^{-}p$ amplitude below the threshold with theoretical uncertainties. This figure indicates that the subthreshold $\bar{K}N$ amplitude is tightly constrained. 

\begin{table}[tbp]	
\begin{center}
\caption{
Kaonic hydrogen level shift $\Delta E$ and width $\Gamma$ by three models in Ref.~\cite{Ikeda:2011pi,Ikeda:2012au} and by SIDDHARTA~\cite{Bazzi:2011zj,Curceanu} .
\label{tab:IHW}
}
\begin{tabular}{lllll}
\hline
 & TW & TWB & NLO & Experiment~\cite{Bazzi:2011zj,Curceanu}  \\ \hline
$\Delta E$ [eV] & 373 & 377 & 306 & $283\pm 36\pm 6$ \\
\vspace{0.2cm}
$\Gamma$ [eV]   & 495 & 514 & 591 & $541\pm 89\pm 22$ \\ 
$\chi^{2}$/d.o.f. & 1.12 & 1.15 & 0.96 &   \\ 
\hline
\end{tabular}
\end{center}
\end{table}

It is important to pin down the source of remaining uncertainties of the amplitude for future investigations. For this purpose, we calculate the $K^{-}n$ scattering in the charge $Q=-1$ sector using the parameters determined in the $Q=0$ sector. The scattering lengths of the $K^{-}p$ and $K^{-}n$ channels with uncertainty regions are given by
\begin{align}
a(K^-n) &=a(I=1)+\dots= 0.57^{+0.04}_{-0.21}+  0.72^{+0.26}_{-0.41}i \quad \textrm{fm},
\end{align}
where dots represent the isospin breaking corrections. By comparing the magnitude of the uncertainty bands of the imaginary parts with that of $a(K^-p)=[a(I=0)+a(I=1)]/2= -0.65\pm 0.10+ ( 0.81\pm 0.15)i \textrm{ fm}$, we find that the constraint on the $I=1$ component is relatively loose, although the sum of $I=0$ and $I=1$ components is strongly constrained by SIDDHARTA. The $K^{-}n$ scattering length is in principle related to the level of kaonic deuterium, which is a challenging subject both in theory and experiment.

Recently, several other groups analyze the SIDDHARTA data with NLO chiral interactions~\cite{Cieply:2011nq,Mai:2012dt,Guo:2012vv,Mizutani:2012gy}. In Table~\ref{tab:comparison}, we summarize the differences of the theoretical treatments (Ref.~\cite{Mizutani:2012gy} does not show the pole position and is omitted from the table). The second column of this table ($V$) indicates the choice of the interaction kernel; ``on-shell'' models use the on-shell factorized kernel, while Ref.~\cite{Mai:2012dt} includes the off-shell effect. Ref.~\cite{Cieply:2011nq} utilizes the separable interaction kernel. The third column ($G$) represents the regularization method of the loop integral. Shown in the fourth column ($f$) is the treatment of the meson decay constant, whether physical values $f_{\pi}\neq f_{K}\neq f_{\eta}$ are used or a common value $f_{\pi}= f_{K}= f_{\eta}$ is adopted. Fifth column classifies the treatment of the $\pi\Sigma$ invariant mass spectrum; ``checked'' models compare the spectrum with the predictions after fitting. In Ref.~\cite{Guo:2012vv} the $\pi\Sigma$ spectrum is used in the fitting with linear combination of different initial states. Sixth column ($\Delta E$, $\Gamma$) shows the method to relate the meson--baryon amplitude with the kaonic hydrogen measurement. Ref.~\cite{Cieply:2011nq} solves the $K^{-}p$ bound state problem numerically, while the other models use the model-independent relation~\eqref{eq:MRR}. In addition, there are some more differences, such as the inclusion of the higher energy coupled channels ($\eta \Lambda$, $\eta \Sigma$, $K\Xi$), constraint for the low energy constants by the $\pi N$ sigma term, definition of $\chi^{2}$, and so on.

\begin{table}[tbp]	
\begin{center}
\caption{
Comparison of models for $S=-1$ meson--baryon scattering in next-to-leading order chiral coupled-channel approach with the SIDDHARTA constraint. See text for the explanation of each column.
\label{tab:comparison}
}
\begin{tabular}{lllllll}
\hline
Model & $V$ & $G$ & $f$ & $\pi\Sigma$ spectrum & $\Delta E$, $\Gamma$ & Pole positions \\ \hline
\cite{Ikeda:2011pi,Ikeda:2012au} NLO & on-shell & dimensional & physical & checked & Eq.~\eqref{eq:MRR} & $1424-26i$, $1381-\phantom{0}81i$ \\
\cite{Cieply:2011nq} NLO30 & separable & form factor & physical & checked & solved & $1418-44i$, $1355-\phantom{0}86i$ \\ 
\cite{Mai:2012dt} & off-shell & dimensional & physical & - & Eq.~\eqref{eq:MRR} & $1428-\phantom{0}8i$, $1467-\phantom{0}75i$ \\ 
\cite{Guo:2012vv} Fit I & on-shell & dimensional & common & used in fit & Eq.~\eqref{eq:MRR} & $1417-24i$, $1436-126i$ \\ 
\cite{Guo:2012vv} Fit II & on-shell & dimensional & physical & used in fit & Eq.~\eqref{eq:MRR} & $1421-19i$, $1388-114i$ \\ 
\hline
\end{tabular}
\end{center}
\end{table}

Although details of models and fitting schemes are different from each other, it turns out that the observed data (cross sections, branching ratios, and SIDDHARTA) are equally well described in all models, which demonstrates the reliability of the chiral coupled-channel approach. On the other hand, some deviations are seen in the subthreshold extrapolations. In Table~\ref{tab:comparison}, we also show the predictions of the $\Lambda(1405)$ pole positions. We find that the first pole is more or less found around 1420 MeV, while the distribution of the second pole quantifies the uncertainty of the extrapolations. At present, the uncertainty may be attributed to the difference of the model setup (such as the treatment of the meson decay constant; see discussion in Ref.~\cite{Guo:2012vv}), but the description of the observed $\pi\Sigma$ spectra, especially that in Ref.~\cite{Agakishiev:2012xk} whose peak locates even below 1400 MeV, seems to be difficult for the models without the lower energy second pole. In any event, the only constraint on the subthreshold $\bar{K}N$ amplitude can be indirectly provided by the $\pi\Sigma$ mass spectra in production experiments. Thus, one of the urgent problems in near future studies is to analyze the $\pi\Sigma$ spectra in recent experiments in Table~\ref{tab:exp} with detailed study of the production mechanisms.

%------------------------------
\subsection{Applications to nuclear systems}
\label{sec:applications}

Study of $\bar{K}$ in nuclei is a hot topic in recent strangeness nuclear physics~\cite{Hyodo:2012pn,Gal}. Among others, the few-nucleon systems with $\bar{K}$ are closely related to the two-body $\bar{K}N$ interaction, and have been studied by various approaches~\cite{Shevchenko:2006xy,Shevchenko:2007zz,Ikeda:2007nz,Yamazaki:2007cs,Dote:2008in,Dote:2008hw,Ikeda:2010tk,Oset:2012gi,Bayar:2012rk,Barnea:2012qa}. All the calculations show that there is a quasi-bound state in the spin $J=0$ $\bar{K}NN$ system. On the other hand, quantitative predictions of the binding energy and decay width have not converged, mainly because of the uncertainty of the subthreshold extrapolation of the $\bar{K}N$ amplitude. Since most of these calculations have been performed before the SIDDHARTA experiment, it is now important to examine how the SIDDHARTA result constrain the properties of the $\bar{K}NN$ quasi-bound state. The well calibrated two-body $\bar{K}N$ amplitude is also an important building block to construct $\bar{K}$ optical potential~\cite{Friedman:2012qy}. In this way, we are now in a position to apply the $\bar{K}N$ scattering models constrained by the SIDDHARTA measurement to various $\bar{K}$-nuclear systems.

%%%%%%%%%%%%%%%%%%%%%%%%%%%%%%%%%%%%%%%%%%%%%%%%%%%%
\section{Concluding remarks}
\label{sec:conclusion}
%%%%%%%%%%%%%%%%%%%%%%%%%%%%%%%%%%%%%%%%%%%%%%%%%%%%

We discuss the $\bar{K}N$ dynamics and the $\Lambda(1405)$ resonance from theoretical and phenomenological viewpoints. Using chiral coupled-channel approach, we establish the $\bar{K}N$ quasi-bound picture of the $\Lambda(1405)$ resonance in Section 2, and we show that the SIDDHARTA measurement provides a strong constraint on the threshold $\bar{K}N$ amplitude. It is emphasized that the new data from recent experiments is crucial to enhance our understanding of the strong dynamics of the $\bar{K}N$ system. Based on the discussions given above, what we need to perform in future studies are
\begin{enumerate}
\item clear definition of the structure of hadron resonances,
\item determination of the $I=1$ component of the $\bar{K}N$ scattering length,
\item detailed study of the $\Lambda(1405)$ production processes, and
\item application of the SIDDHARTA-constrained models to $\bar{K}$-nuclear systems.
\end{enumerate}
The first point is important to discuss the structure of $\Lambda(1405)$ from  a broader standpoint. For further constraints on the $\bar{K}N$ interaction, the second and third are required to complement the precision of the threshold $\bar{K}N$ amplitude and to extract the information of subthreshold $\bar{K}N$ amplitude from experimental $\pi\Sigma$ spectra, respectively. The last one is to provide more reliable predictions for the possible $\bar{K}$ bound states in nuclear systems. We expect that this field will be further developed through the mutual collaboration between theory and experiments.

%%%%%%%%%%%%%%%%%%%%%%%%%%%%%%%%%%%%%%%%%%%%%%%%%%%%
\section*{Acknowledgments}
%%%%%%%%%%%%%%%%%%%%%%%%%%%%%%%%%%%%%%%%%%%%%%%%%%%%

The author is grateful to the collaborators of a series of works presented here, especially to Yoichi Ikeda, Daisuke Jido and Wolfram Weise. He also thanks the organizers for the kind invitation to the fruitful conference.
This work is partly supported by the Grant-in-Aid for Scientific Research from MEXT and JSPS (No. 24105702 and No. 24740152)
and the Global Center of Excellence Program by MEXT, Japan, through the Nanoscience and Quantum Physics Project of the Tokyo Institute of Technology. 



%% References


%\bibliographystyle{elsarticle-num}
%%\bibliography{<your-bib-database>}
%\bibliography{refs,refs05,myrefs}


%%% For references without a BibTeX database:
%
%\begin{thebibliography}{00}
%
%%% \bibitem must have the following form:
%%%   \bibitem{key}...
%%%
%
%\bibitem{article} F. Ransome, Nucl.\ Phys.\ A007 (2010) 234.
%
%\bibitem{arxiv}  A. Einstein et al., hep-lat/69999.
%
%\bibitem{book} M. Curie and I. Joliot-Curie, Pierre Curie Book, Benjamin, New York, 1900.
%
%\end{thebibliography}

\end{document}